\documentclass[prb,twocolumn,showpacs,amsmath,amssymb,superscriptaddress,groupedaddress]{revtex4-1}

\usepackage{graphicx}
\usepackage{bm}
\usepackage{physics}
\usepackage{multirow}
\usepackage{hyperref}
\hypersetup{pdfnewwindow=true, colorlinks=true,
 linkcolor=blue, anchorcolor=blue,
 citecolor=blue, filecolor=blue,
 menucolor=blue, urlcolor=blue}

\usepackage{tabularx}
\newcolumntype{C}[1]{>{\centering\arraybackslash}p{#1}}

\begin{document}

\title{Classification of materials with phonon angular momentum and microscopic origin of angular momentum}

\author{Sinisa Coh}
\affiliation{Materials Science and Mechanical Engineering, University of California Riverside, CA 92521, USA}

\date{\today}

\begin{abstract}
We group materials into five symmetry classes and determine in which of these classes phonons carry angular momentum in the Brillouin zone, away from a high-symmetry point, line, or plane.  In some materials phonons acquire angular momentum via the forces induced by relative displacements of atoms out of their equilibrium positions.  However, for other materials, such as ferromagnetic iron, phonon angular momentum arises from the forces induced by relative velocities of atoms.  These effects are driven by the spin-orbit interaction.
\end{abstract}

\maketitle

A phonon is a quantum of ionic motion in a solid and is characterized by a branch index $\nu$, a crystal momentum $\bm q$, and frequency $\omega$.  Phonons can also carry angular momentum $\bm l$.\cite{McLellan1988, Zhang2014, Zhang2015, Zhu2018}  Such phonons are characterized by circular, or elliptical, motion of ions. As shown in Ref.~\onlinecite{Zhang2014}, each phonon degree of freedom contributes to the energy of the solid by $$\hbar \omega (n + 1/2)$$ and to the total angular momentum of the solid, $L$, by $${\bm l} \left( n + 1/2 \right).$$  Here $n$ is the Bose-Einstein occupation factor. The magnitude of $l^z$ for a circularly polarized phonon mode moving in the $xy$-plane is $\pm \hbar$, while for the elliptically polarized mode it can have any value between $-\hbar$ and $\hbar$.\footnote{On general quantum mechanical principles, for any rotationally invariant solid one can find a basis of eigenstates with an expectation value of angular momentum that is an integer multiple of $\hbar$. However, this is not the case for a solid with periodic boundary condition, as periodic boundary condition breaks the rotational symmetry of the system.}  In the zero-temperature limit ($n=0$) each circularly polarized phonon contributes to the angular momentum by one half of $\pm \hbar$, similar to how each phonon contributes to the energy with one half of $\hbar \omega$.  At elevated temperature, each excited circularly polarized phonon mode contributes $\pm \hbar$ to the total angular momentum, and $\hbar \omega$ to the total energy.

Here we discuss which classes of materials can have phonons with non-zero angular momentum $\bm l$ at generic non-symmetric part of the Brillouin zone. Next, for each material class, we discuss the microscopic origin of the angular momentum in the lattice.  While for some materials the angular momentum originates within the Born-Oppenheimer approximation,\cite{Born1954} in other materials, such as ferromagnetic iron, the phonon angular momentum is acquired only by going beyond that approximation. We present first-principles calculations for both cases.

Phonon angular momentum and the underlying forces that are responsible for its microscopic origin play a crucial role in the diverse range of effect ranging from the phonon Hall effect,\cite{Strohm2005, Grissonnanche2020, Park2020, Flebus2022} magnetic moment of a phonon,\cite{Juraschek2017,Juraschek2019, Park2020, Ren2021, Xiong2022} Einstein de-Haas effect,\cite{Zhang2014,Garanin2015,Nakane2018, Tauchert2022} topological phononic insulators,\cite{Kane2014}, Dirac materials,\cite{Hu2021} driven chiral phonons\cite{Juraschek2020, Geilhufe2021, Geilhufe2023}, and other effects.\cite{Hamada2018}

\section{Symmetry}
\label{sec:symmetry}

\begin{table*}[!t]
\caption{\label{tab:summary} Angular moment of a phonon at a generic non-symmetric point (away from high-symmetry points, lines, and planes) in the Brillouin zone in five classes of materials discussed in the text.  Angular momentum is allowed in classes III, IV, and V.}
\begin{ruledtabular}
\begin{tabular}{c C{0.7cm} C{0.7cm} C{0.7cm} ccc}
\multirow{2}{*}{Class} & \multicolumn{3}{c}{Present symmetries} 
& \multirow{2}{*}{Angular momentum?}
& Microscopic origin of & \multirow{2}{*}{Examples}  \\
\cline{2-4}
& $\cal P$ & $\cal T$ & ${\cal PT}$ & & angular momentum & \\
\hline
I   & \checkmark & \checkmark & \checkmark & No  &                                                          & Si, Au       \\
II  & $\cross$   & $\cross$   & \checkmark & No  &                                                          & Cr$_2$O$_3$  \\
III & $\cross$   & \checkmark & $\cross$   & Yes & Force-constant matrix $F^{\alpha \beta}_{ij}$            & WC, WSe$_2$, HgS, SiO$_2$, Te  \\
IV  & \checkmark & $\cross$   & $\cross$   & Yes & Velocity-force matrix  $G^{\alpha \beta}_{ij}$           & Fe, Ni, Co   \\
V   & $\cross$   & $\cross$   & $\cross$   & Yes & Both $F^{\alpha \beta}_{ij}$ and $G^{\alpha \beta}_{ij}$ & MnGe         \\
\end{tabular}
\end{ruledtabular}
\end{table*}

We start by using the symmetry arguments to determine in which materials phonons have angular momentum at a generic non-symmetric part of the Brillouin zone, away from the high-symmetry points, lines, and planes. Restriction to generic non-symmetric part of the Brillouin zone greatly simplifies the analysis, as it allows us to only consider the point-group symmetries of the system.  Analysis that includes high-symmetry parts of the Brillouin zone would necessarily have to consider the irreducible representation of the magnetic space group associated with each phonon, as at high-symmetry points only modes associated with a specific representation would have phonon angular momentum.  Such an analysis is beyond the scope of this manuscript.  Second reason for focusing on generic non-symmetric points is that many physical effects, such as Einstein de-Haas effect for example, rely on changes in population of phonons across the entire Brillouin zone. Therefore, in such processes high-symmetry points, lines, or planes are irrelevant as they occupy part of the phonon's Brillouin zone with no volume, so virtually no phonons correspond to those parts of the Brillouin zone. There are, of course, many other important situations where the high-symmetry points are the only relevant parts of the Brillouin zone.  One example is the lowest order Raman effect,\cite{Tatsumi2018} where one approximates that the relevant phonons are at a high symmetry point (Brillouin zone origin).

In what follows, when we say that a material has a time-reversal symmetry, we mean that the time-reversal symmetry is contained in the point group of the material.  Therefore, we allow for the possibility that in some materials (for example, some anti-ferromagnets) the actual space group symmetry element consists of time-reversal symmetry followed by a fractional translation of the lattice.

Under these assumptions, we now perform the symmetry analysis of the phonon angular momentum in a material.  Inversion operation ($\cal{P}$) of the crystal transforms a phonon (at a generic non-symmetric point) with angular momentum $\bm{l}$ and linear momentum $\bm{q}$ into a phonon with the same angular momentum but opposite linear momentum,
\begin{align}
  {\cal P} : \  (\bm{l},\bm{q}) \longrightarrow (\bm{l},-\bm{q}).
  \label{eq:sym_p}
\end{align}
Similarly, for the time-reversal operation ($\cal{T}$) we have,
\begin{align}
{\cal T} : \  (\bm{l},\bm{q}) \longrightarrow (-\bm{l},-\bm{q}).  
  \label{eq:sym_t}
\end{align}
Therefore, for an operation $\cal{P T}$ that consists of time reversal followed by a spatial inversion, we have,
\begin{align}
{\cal PT} : \  (\bm{l},\bm{q}) \longrightarrow (-\bm{l},\bm{q}).
\label{eq:pt}
\end{align}
The product $\cal{P T}$ is the only operation of the crystal that leaves $\bm{q}$ unchanged at a generic non-symmetric point of the Brillouin zone. Therefore, from Eq.~\eqref{eq:pt} it follows that if and only if ${\cal PT}$ is a symmetry, a phonon at such $\bm{q}$ will have a zero angular momentum.

As mentioned earlier, in the derivation above, we are considering $\bm q$-points away from high-symmetry points, lines, or planes of the Brillouin zone.  This is a crucial assumption here, as this guarantees that the phonon under consideration is non-degenerate. Otherwise, if we had a pair of degenerate modes, then the operations $\cal{P}$, $\cal{T}$, or $\cal{P T}$ could transform one degenerate phonon mode into another and the analysis above would not hold. Therefore, one can construct a phonon with angular momentum at high-symmetry points in a much wider range of materials, even those with $\cal{P T}$ symmetry. As a simple example, graphene\cite{Tatsumi2018} has degenerate phonons at $\bm{q}=\bm{0}$ and one can find a basis in the degenerate subspace of phonons in which individual phonon modes have non-zero angular momentum.  Furthermore, it is worth mentioning that in any system, regardless of its symmetries, one can construct a pattern of atomic displacements with angular momentum by taking a linear combination of phonons at $\bm q$ and $-\bm q$.  In this work we are specifically focusing only on atomic displacements (phonons) that are characterized by a single $\bm q$-point in the Brillouin zone.

In Table~\ref{tab:summary} we summarized five classes of materials with respect to the presence of inversion symmetry ($\cal P$), time-reversal symmetry ($\cal T$), or their combination (${\cal PT}$).

Materials in class I are defined as having all three symmetries: $\cal P$, $\cal T$, and ${\cal PT}$.  Since these materials have ${\cal PT}$ as one of their symmetries, they can't have a phonon angular momentum at a generic non-symmetric point in the Brillouin zone.  For materials in class II $\cal P$ and $\cal T$ are not symmetries, but their product ${\cal PT}$ is a symmetry, which leads to the same conclusion.  Materials in class III have broken $\cal P$ and ${\cal PT}$ while $\cal T$ itself is a symmetry.  On the other hand, in class IV has broken $\cal T$ and ${\cal PT}$ while $\cal P$ is a symmetry.  In class V all three symmetries are broken.

\section{Phonon equation of motion}

The dynamics of ions is typically described\cite{Born1954} within the lowest order Born-Oppenheimer approximation via the force-constant matrix $F^{\alpha \beta}_{ij}$. This matrix describes force induced on the $i$-th atom (in direction $\alpha$) by a displacement of the $j$-th atom (in direction $\beta$).  Therefore, equation of motion for the $i$-th nucleus (with mass $M_i$) is given by,
\begin{align}
- M_i \frac{d^2 x^{\alpha}_i}{dt^2} = \sum_{j\beta} F^{\alpha \beta}_{ij} x^{\beta}_j.
\end{align}
This description is only approximate, as the true dynamics of nuclei is quantum-mechanical, and it therefore can't be fully described by a classical equation of motion. Within the Born-Oppenheimer approximation the total wavefunction $\Psi({\bm x}, {\bm x}^{\rm elec} )$ for nuclei and electrons is approximated as $$\Psi({\bm x}, {\bm x}^{\rm elec}) = \psi({\bm x}) \phi_{{\bm x}} ({\bm x}^{\rm elec}),$$ where $\phi$ is the instantaneous electronic wavefunction parameterized by fixed location of nuclei.  Using this ansatz within the full Hamiltonian for nuclei and electrons results in an effective Schrodinger equation for nuclear wavefunction $\psi({\bm x})$.  The derivatives of electronic eigenenergies $\partial_{\bm x} E_{\bm x}$ associated with electronic wavefunction $\phi$ give an effective scalar potential for ions, while the $\langle \phi_{\bm x} \vert i \nabla_{\bm x}  \vert \phi_{\bm x} \rangle$ acts as an effective vector potential.\cite{Moody1986,Abedi2010,Tao2012}

\begin{table*}[!t]
\caption{\label{tab:berry}Relationship between various Berry-like objects that depend on either electronic Berry-connection ${\cal A}_k$ in the reciprocal space ($k$) or on the ionic Berry-connection ${\cal A}_r$ in the space of ionic positions ($r$).} 
\begin{ruledtabular}
\begin{tabular}{c|l|l|l}
Dimensionality & Electronic Berry-like terms & Mixed electronic-ionic Berry-like terms & Ionic Berry-like terms \\
\hline
\multirow{2}{*}{1} & Polarization     &  Born effective charge                 & \\
  & $P = \int {\cal A}_k dk$    &  $ Z = \int \partial_r {\cal A}_k dk$  & \\
\hline
\multirow{2}{*}{2} & Anomalous Hall conductivity              &                           & Velocity-force  \\
  & $\sigma = \int \partial_k {\cal A}_k dk$             &               & $G = \int \partial_r {\cal A}_r dk$  \\
\hline
\multirow{2}{*}{3} & (component of) magnetoelectric coupling              &               & \\
  & $\theta = \int {\cal A}_k \partial_k {\cal A}_k dk$                  &  & \\
\end{tabular}
\end{ruledtabular}
\end{table*}

First we consider the effective scalar potential experienced by nuclei.  Expanding around the ground state in terms of small atomic displacements, one obtains the force-constant matrix $F^{\alpha \beta}_{ij}$.  Such a force-constant matrix can be computed by first solving a family of electronic Schrodinger equations as a function of all ionic coordinates ${\bm x}$
\begin{align}
H_{\bm x} \phi_{\bm x} = E_{\bm x} \phi_{\bm x},
\end{align}
and then taking the second derivative of $E_{\bm x}$ with respect to atomic coordinates,
\begin{align}
F^{\alpha \beta}_{ij} = \frac{\partial^2 E_{{\bm x}}}{\partial x^{\alpha}_i \partial x^{\beta}_j}.
\label{eq:fij}
\end{align}
This approximation can be improved by considering the effective vector potential, as well as higher orders in the expansion (of either scalar or vector potential).  In such an expansion the higher order terms in atomic positions $x_j$ as well as their time derivatives can occur.  In the lowest order the dynamics is then described by an infinite series of additional terms,
\begin{align}
-M_i \frac{d^2 x^{\alpha}_i}{dt^2} = & \sum_{j\beta} F^{\alpha \beta}_{ij} x^{\beta}_j + \notag \\
+ & \sum_{j\beta} G^{\alpha \beta}_{ij} \frac{d x^{\beta}_j}{dt} + \notag \\
+ & \sum_{j k \beta \gamma} H^{\alpha \beta \gamma}_{ijk} x^{\beta}_j x^{\gamma}_k + ...
\label{eq:eom}
\end{align}
In contrast to the force-constant matrix $F^{\alpha \beta}_{ij}$, the velocity-force constant matrix $G^{\alpha \beta}_{ij}$ can't be computed from the energies of the electronic Schrodinger equation, but is instead computed from the effective vector potential, which results in,
\begin{align}
G^{\alpha \beta}_{ij} = 2 \hbar {\rm Im} \bigg\langle \frac{\partial \phi_{{\bm x}}}{\partial x^{\alpha}_i} \bigg| 
\frac{\partial \phi_{{\bm x}}}{\partial x^{\beta}_j} \bigg\rangle.
\label{eq:gij}
\end{align}
While this term appears in the so-called Born-Huang approximation\cite{Born1954} it is often ignored as it is assumed to be small.  The term $G$ also appears in the quantum-mechanical description of the motion of three identical nuclei on a triangle\cite{mead1979} in an external magnetic field and is seen as an extension of the Aharonov-Bohm effect.\cite{Mead1980} In a semi-classical description of atomic motion, as in Eq.~\ref{eq:eom} the term proportional to $G$ can be thought of as a Lorentz force on a charged particle in an effective magnetic field.\cite{Resta2000} Another example where Eq.~\ref{eq:gij} appears in the literature is in the phonon Hall effect, that was first discovered experimentally\cite{Strohm2005,Inyushkin2007} and then assigned the theoretical origin.\cite{Sheng2006}  In this context the literature refers to the Eq.~\ref{eq:gij} as the Raman spin-phonon interaction. We refer the reader to references in Ref.~\onlinecite{Zhang2011} for a history of the phonon Hall effect.

We incorporate Eq.~\ref{eq:gij} into equation of motion by using ansatz
\begin{align}
  x_i^{\alpha} (t) = A M_i^{-1/2} {\rm Re} \left( \xi^{\alpha}_i e^{i \omega t} \right).
  \label{eq:ansatz}
\end{align}
Here $A$ is an arbitrary real constant to be determined later.  The phonon eigenvectors ($\xi^{\alpha}_i$, defined to be normalized to unity) and frequencies ($\omega$) are then obtained by diagonalizing a generalized eigenvalue problem,
\begin{align}
  \sum_{j \beta} \frac{1}{\sqrt{M_i M_j}} \left( F_{ij}^{\alpha \beta} + i \omega G_{ij}^{\alpha \beta} \right) \xi^{\beta}_j = \omega^2 \xi^{\alpha}_i.
  \label{eq:eigval}
\end{align}
Clearly, if $G=0$ then the equation above reduces to the well known problem of finding eigenvalues of a dynamical matrix $F_{ij}^{\alpha
\beta}/\sqrt{M_i M_j}$. With $G \neq 0$ the extra term proportional to $G$ has the form of a frequency-dependent correction to the force-constant matrix.

We briefly comment on the mathematical form of Eq.~\ref{eq:eom} as a Berry curvature in the space of atomic coordinates.\cite{Resta2000}  In contrast, Berry curvature in the reciprocal space (for fixed atomic coordinates) is related to the off-diagonal $\sigma_{xy}$ conductivity, appearing in the context of the anomalous Hall effect and integer quantum Hall effect.\cite{Sundaram1999,Haldane2004}  The relationship between different Berry-like quantities is shown in Table~\ref{tab:berry}.

\section{Microscopic origin of angular momentum}

Given the equation of motion Eq.~\ref{eq:eom} we now come to the question of the origin of the phonon angular momentum. 

\subsection{Class I and III}

In class I all three symmetries are present (${\cal P}$, ${\cal T}$, and ${\cal PT}$) so both $F^{\alpha \beta}_{ij}$ and $G^{\alpha \beta}_{ij}$ terms in Eq.~\ref{eq:eom} preserve these symmetries and phonons don't have angular momentum at a generic non-symmetric point in the Brillouin zone.  

In class III the time-reversal ${\cal T}$ is a symmetry, but the inversion symmetry is broken. Therefore trivially, the inversion symmetry breaking will spill into the force-constant matrix $F^{\alpha \beta}_{ij}$.  This can easily be demonstrated on a toy example shown in Fig.~\ref{fig:summ}. (An example of a real solid with broken inversion symmetry is discussed in Ref.~\onlinecite{Moseni2022}.)  The unit cell there consists of two atoms, indicated with green and orange spheres.  If the green atom is exactly in the center of the square formed by orange spheres (left panel), we have inversion symmetry, and therefore force-constants (springs) $F^{\alpha \beta}_{ij}$ between green and four orange atoms will all be equal by symmetry.  However, if we break the inversion symmetry by displacing the green atom away from the center (middle panel), then two top force-constants $F^{\alpha \beta}_{ij}$ will have a different value (thick black line) than the two bottom force-constants (thin black line).  The fact that $F^{\alpha \beta}_{ij}$ has explicitly broken inversion symmetry means, by our earlier symmetry analysis from Sec.~\ref{sec:symmetry}, that if we solve equation of motion Eq.~\ref{eq:eom} with such $F^{\alpha \beta}_{ij}$ that the resulting phonon eigenvectors will have angular momentum at a generic non-symmetric point in the Brillouin zone. This mechanism, therefore, will be the source of the phonon angular momentum in classes III (and partially in class V as well).

In class III the phonon angular momentum of the same phonon branch at $\bm{q}$ must have opposite sign to that at $-\bm{q}$.  This follows from Eq.~\ref{eq:sym_p} and such phonon band structure is sketched in Fig.~\ref{fig:summ}.

\subsection{Class IV}

The situation in class IV is somewhat more complex. In this class of materials, containing ferromagnets such as iron, the inversion symmetry is preserved while the time-reversal symmetry is broken.  As a general principle, one would expect that time-reversal breaking in the electronic subsystem must somehow spill into the ionic subsystem, as electrons and ions are coupled. While this is true, the spilling of the time-reversal breaking into ions does not occur in the first term of the expansion in Eq.~\ref{eq:eom}, the force constant matrix $F^{\alpha \beta}_{ij}$, but it does spill into the velocity-force matrix $G^{\alpha \beta}_{ij}$. Let us demonstrate this from the definition of $F^{\alpha \beta}_{ij}$ and $G^{\alpha \beta}_{ij}$.

The force-constant $F^{\alpha \beta}_{ij}$ was defined in Eq.~\ref{eq:fij} as the second derivative of the total energy $E_{\bm x}$ with respect to atom coordinate.  For solid of any symmetry, the total energy $E_{\bm x}$ is a scalar that is invariant under the time-reversal operation,
\begin{align}
  {\cal T} : \  E_{\bm x} \longrightarrow E_{\bm x}
  \label{eq:sym_E}
\end{align}
for any set of atom coordinates $\bm x$.  Therefore, the derivatives of $E_{\bm x}$ with respect to $\bm x$ are also unchanged under ${\cal T}$, so the force-constant matrix is unchanged as well,
\begin{align}
  {\cal T} : \  F_{ij}^{\alpha \beta}  \longrightarrow F_{ij}^{\alpha \beta}.
  \label{eq:sym_F}
\end{align}
This holds regardless of whether the solid itself is in a time-reversal symmetric ground state or not.  Therefore,  Eq.~\ref{eq:sym_F} holds even in a ferromagnet like bulk Fe, or any other materials in class IV.  For example, ferromagnetic bulk Fe magnetized along the positive $\hat{\bm z}$ direction will have exactly the same force-constant matrix $F^{\alpha \beta}_{ij}$ as its time-reversed image where the magnetic moment is reversed to point along the negative $\hat{\bm z}$ axis. This holds true even if spin-orbit interaction, or any other relativistic effect, is included in the calculation.

\begin{figure}[!t]
\centering
\includegraphics[width=3.2in]{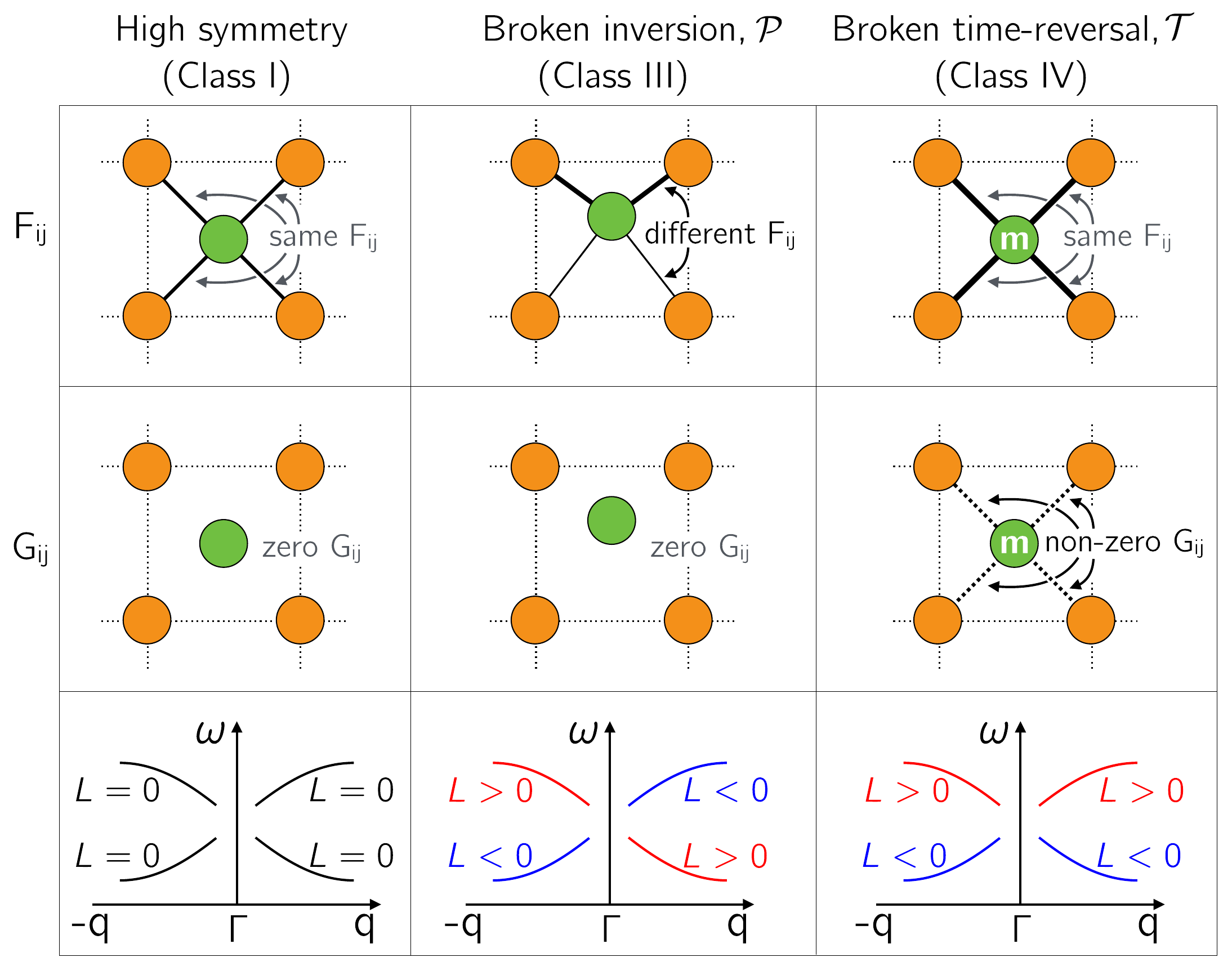}
\caption{\label{fig:summ} Sketch of microscopic origins of the phonon angular momentum in a fictitious material with two atoms per unit cell (drawn as green and orange circles).  The first column corresponds to material where $\cal P$, $\cal T$, and ${\cal PT}$ symmetries are present.  In the second column, inversion ($\cal P$) is broken by displacement of the green atom (class III).  In the third column, time-reversal ($\cal T$) is broken by (class IV) magnetism on the green atom (magnetic moment points out of the page). Strengths of force constant matrices $F^{\alpha \beta}_{ij}$ and velocity-force constant matrices $G^{\alpha \beta}_{ij}$ are indicated with black lines in the first and the second row for all three cases. The third row shows sketches of the corresponding phonon band structures and phonon angular momenta in the vicinity of the Brillouin zone origin.  Signs of angular momenta for two sketched phonon branches follow from Eqs.~\ref{eq:sym_p} and \ref{eq:sym_t}.  Phonon angular momentum exactly at the origin is excluded from the sketch, as in this work we focus on generic non-symmetric points of the Brillouin zone only.  Phonon angular momentum in the second column (broken inversion, $\cal P$, class III) arises from the asymmetric force-constants $F_{ij}$ (as sketched in the top panel of the second column). Analogously, in the third column (broken time-reversal, $\cal T$, class IV) phonon angular momentum arises from the presence of non-zero $G_{ij}$ (as sketched in the middle panel of the third column).}
\end{figure}

Since the force-constant matrix $F^{\alpha \beta}_{ij}$ is unaware of the time-reversal symmetry breaking in the solid, any ionic motion in the class IV material that is driven only by the force-constant matrix $F^{\alpha \beta}_{ij}$ will preserve time-reversal symmetry.  Therefore, following symmetry analysis from Sec.~\ref{sec:symmetry}, the phonons in class IV material described only by $F^{\alpha \beta}_{ij}$ will not have angular momentum at a generic non-symmetric point in the Brillouin zone. This same observation can be again made on our toy model from Fig.~\ref{fig:summ}. Imagine that instead of displacing the green
atom, we make the green atom magnetic, therefore breaking the
time-reversal symmetry in the solid.  Magnetization of the green atom
is pointing out of the page.  Since the distribution of charge on the
atom was changed when we made the atom magnetic, one might expect that the resulting force constants between the green and orange atoms will change as well.  And they do, but clearly, the changes to all four force constants must be equal.  More importantly, these four force constants would change by the same amount, regardless of whether the magnetic moment on the green atom in Fig.~\ref{fig:summ} is pointing in or out of the page.  Therefore, adding a magnetic moment to the green atom did not change the symmetry in the force constant matrices $F^{\alpha \beta}_{ij}$, so the underlying phonons did not acquire angular momentum from changes in $F^{\alpha \beta}_{ij}$.

Now let us consider the equation of motion for a solid in class IV that includes the next term in the expansion, the velocity-force constant term $G^{\alpha \beta}_{ij}$. While force-constant matrix is defined in terms of a scalar quantity (total energy $E_{\bm x}$) which doesn't change under time-reversal operation, the velocity-force constant is defined in terms of the electron wavefunctions (see Eq.~\ref{eq:gij}) which do change under time-reversal.  The time-reversal operation, when acting on electron wavefunctions, is represented by an anti-unitary operator $\cal \hat{T}$.  For any two wavefunctions $\ket{\phi_1}$ and $\ket{\phi_2}$ we define $\ket{\phi'_1} = {\cal \hat{T}} \ket{\phi_1}$ and $\ket{\phi'_2} = {\cal \hat{T}} \ket{\phi_2}$.  For any anti-unitary operation, by definition, we have
$\braket{\phi'_1}{\phi'_2} =  \left[ \braket{\phi_1}{\phi_2}    \right] ^*$.
This gives us,
\begin{align}
{\cal T} : \ 
\bigg\langle \frac{\partial
\phi_{{\bm x}}}{\partial x^{\alpha}_i} \bigg|
\frac{\partial \phi_{{\bm x}}}{\partial x^{\beta}_j} \bigg\rangle
\rightarrow
\left[ \bigg\langle \frac{\partial
\phi_{{\bm x}}}{\partial x^{\alpha}_i} \bigg|
\frac{\partial \phi_{{\bm x}}}{\partial x^{\beta}_j} \bigg\rangle \right]^*.
\end{align}
Since $G^{\alpha \beta}_{ij}$ from Eq.~\ref{eq:gij} depends on the imaginary part of this overlap, we conclude that $G_{ij}$ changes sign under time-reversal operation,
\begin{align}
  {\cal T} : \  G_{ij}^{\alpha \beta} \longrightarrow - G_{ij}^{\alpha \beta}.
  \label{eq:sym_G}
\end{align}
In other words, if a solid has a broken time-reversal symmetry, that breaking will spill into the velocity-force constant matrix $G_{ij}^{\alpha \beta}$, and therefore ionic motion will also experience broken time-reversal symmetry.  For example, this means that reversing the direction of the magnetization in bulk ferromagnetic iron from $+\hat{\bm z}$ to $-\hat{\bm z}$ will change the sign of $G_{ij}^{\alpha \beta}$ as well.

(Alternatively, we can demonstrate this point also by considering the expansion in Eq.~\ref{eq:eom}. From this expansion, the $G^{\alpha \beta}_{ij}$ term can be seen as the second derivative of energy, once with respect to atomic position  and once with respect to atomic velocity.  Therefore, velocity-force $G^{\alpha \beta}_{ij}$ changes sign under time-reversal as the velocity also changes sign under time-reversal.  In fact, any term in the equation of motion Eq.~\ref{eq:eom} with an odd number of time derivatives will change sign under time-reversal symmetry.)

Therefore, the phonon angular momentum in ferromagnetic iron, and any other material in class IV, will originate not from $F^{\alpha \beta}_{ij}$ but from terms with an odd number of time-derivatives.  Clearly, the dominant contribution will come from $G^{\alpha \beta}_{ij}$, as it is the lowest term in the expansion with the correct number of time-derivatives.

In contrast to class III, in class IV the phonon angular momentum of the same phonon branch at $\bm{q}$ must have the same sign as that at $-\bm{q}$.  This follows from Eq.~\ref{eq:sym_t} and such phonon band structure is sketched in Fig.~\ref{fig:summ}.

\subsection{Class II}

Somewhat more involved is the case of materials in class II.  These are anti-ferromagnets, such as Cr$_2$O$_3$ where the inversion symmetry operation $\cal P$ is centered in between two magnetic Cr atoms with opposing magnetic moments.  Therefore, while the spatial inversion $\cal P$ is broken (as two Cr atoms with opposing magnetic moments are not equivalent) the product ${\cal PT}$ of spatial inversion with the time-reversal operation is a symmetry.  The presence of  ${\cal PT}$ symmetry ensures that phonons at a generic non-symmetric point in the Brillouin zone don't have phonon angular momentum. However, since time-reversal itself is broken, the velocity-force $G_{ij}^{\alpha \beta}$ is generally non-zero in class II, but this doesn't induce the phonon angular momentum as $G_{ij}^{\alpha \beta}$ itself must be ${\cal PT}$-symmetric.

\subsection{Class V}

Finally, we now discuss class V in which the situation is the simplest from the point of view of symmetry. Now none of the three operations ($\cal P$, $\cal T$, and ${\cal PT}$) are a symmetry of the systems.  Therefore, the phonon angular momentum is now induced both by $F_{ij}^{\alpha \beta}$ and $G_{ij}^{\alpha \beta}$.  Furthermore, since there is no symmetry now that would map generic non-symmetric $\bm{q}$ to $-\bm{q}$, there is now no relationship between phonon frequency and phonon angular momentum at generic non-symmetric $\bm{q}$ to $-\bm{q}$. This is in contrast to the situation in classes III and IV, as sketched in the bottom panels of Fig.~\ref{fig:summ}.

\section{Example: broken inversion-symmetry (class III)}

Now let us consider phonon angular momentum in a few specific materials, as calculated from the first-principles approach.  First, we will consider material from class III with broken inversion symmetry but with time-reversal symmetry. According to our earlier analysis, phonon angular momentum in this material will originate from the inversion symmetry breaking of the force constant matrix $F^{\alpha \beta}_{ij}$.

As an example of material in class III we considered tungsten-carbide, WC. Its structure can be seen as an alternating series of tungsten and carbon hexagonal sheets.  The space group of WC is P$\bar{6}$m2 (number 187).  This space group does not contain inversion symmetry.  The origin of the inversion symmetry breaking is not a displacement of either W or C atoms, it instead originates from the fact that W atom is different from the C atom.  If both W and C sites were populated with the same type of atom, the space group would become P$6_3/$mmc (number 194) which does contain inversion symmetry.

Some other interesting materials in class III are $\alpha$-HgS as studied in Ref.~\onlinecite{Ishito2022}, $\alpha$-SiO$_2$ studied in Ref.~\onlinecite{Ueda2023}, as well as Te studied in Ref.~\onlinecite{Chen2022}.

We computed phonon band structure of WC within the density functional perturbation theory, as implemented in the Quantum-ESPRESSO computer package.\cite{espresso}  We approximate exchange correlation with the Perdew–Burke-Ernzerhof (PBE) approximation.\cite{Perdew_1996}  We compute dynamical matrices on a regular 6x6x6 grid of points and interpolate on a denser grid of points to plot the phonon band structure.

Given a normalized eigenvector $\xi^{\alpha}_i$ of the dynamical matrix (here $\alpha$ is Cartesian direction and $i$ is atomic index) the phonon angular momentum is given as,
\begin{align}
  l^z = 2 \hbar \sum_i {\rm Im} \left( \xi^x_i \bar{\xi}^y_i \right).
  \label{eq:ldef}
\end{align}
Similar expressions hold for phonon angular momentum in $x$ and $y$ directions.  The derivation of Eq.~\ref{eq:ldef} is given in Refs.~\onlinecite{McLellan1988, Zhang2014}. Here we only sketch the derivation of the phonon angular momentum in the semi-classical language. Classical atomic displacements of the lattice vibration are given by Eq.~\ref{eq:ansatz}.  If we choose arbitrary constant $A$ in Eq.~\ref{eq:ansatz} to equal $\sqrt{2 \hbar / \omega}$ then classical energy in the atomic vibration equals $\hbar \omega$, as expected.  With this normalization, the computation of the angular momentum of the lattice vibration is now simply
\begin{align}
  l^{\alpha} = \epsilon^{\alpha \beta \gamma}\sum_{i \beta \gamma} M_i x_i^{\beta} \frac{d x_i^{\gamma}}{d t}.
\end{align}
If we use now Eq.~\ref{eq:ansatz} with $A=\sqrt{2 \hbar / \omega}$ we get Eq.~\ref{eq:ldef}, in agreement with quantum-mechanical derivation from Refs.~\onlinecite{McLellan1988, Zhang2014}. (Here $\epsilon^{\alpha \beta \gamma}$ is the Levi-Civita symbol.  $\epsilon^{\alpha \beta \gamma} =  1$ for any even perturbation of three indices, $\epsilon^{\alpha \beta \gamma} = -1$ for odd perturbation, and $\epsilon^{\alpha \beta \gamma} = 0$ if any two of the indices are repeating.)

\begin{figure}[!t]
\centering
\includegraphics{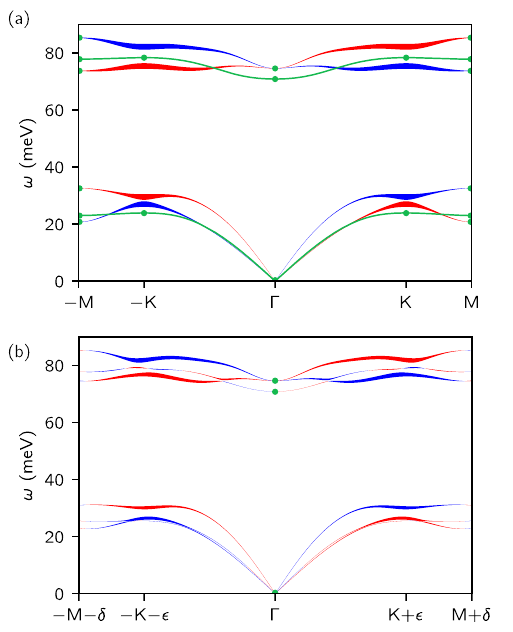}
\caption{\label{fig:wc}  The phonon band structure of tungsten-carbide (class III) along the high-symmetry path (a) in reciprocal space, and along the path that is shifted off the high-symmetry path by around 10\% of the reciprocal lattice vector (b).  Phonons with positive (negative) angular momentum along the $z$-axis are indicated by red (blue) color.  The thickness of the line is proportional to the magnitude of the phonon angular momentum component $l^z$. The maximum line thickness (at the K-points) in the figure corresponds to the circularly polarized phonon mode with $l=\pm \hbar$. Phonons at high-symmetry lines and points that by symmetry have $l^z=0$ are indicated with green line and dot.}
\end{figure}

Figure~\ref{fig:wc} shows calculated phonon band structure, and phonon angular momentum, of tungsten-carbide, WC. Phonons with angular momentum along the $z$-axis are indicated with red and blue color.  Different colors correspond to different signs of the phonon angular momentum $l^z$.  As can be seen from Fig.~\ref{fig:wc} phonons in class III have opposite angular momentum at $\bm{q}$ and $-\bm{q}$, which is in agreement with Eq.~\ref{fig:wc} and sketch in Fig.~\ref{fig:summ}.

Phonons that have $l^z=0$ by symmetry are indicated with green line or circle.  As can be seen from panel (a) of Fig.~\ref{fig:wc} there are two phonon branches, indicated with a green line, that have $l^z=0$ along the entire $\Gamma$--$K$--$M$ path.  Furthermore, all phonons have $l^z=0$ at inversion symmetry invariant $M$ point, but only two phonons have $l^z=0$ at the $K$ point.  At the $\Gamma$ point one can choose a basis of phonon modes so that $l^z=0$ for all phonon branches.  However, for the doubly degenerate modes at the $\Gamma$ point one can choose a basis so that $l^z \neq 0$. Panel (b) of Fig.~\ref{fig:wc} shows phonon band structure on a straight line in phonon Brillouin zone which passes only near $K$ and $M$ points, but never goes through $K$ and $M$.  This path is chosen so that it avoids high-symmetry points, lines, and planes (except for origin, $\Gamma$).  As can be seen from panel (b), all six phonons now have non-zero $l^z$ along the entire path (except for $\Gamma$), in agreement with the discussion in Sec.~\ref{sec:symmetry}.

\section{Example: broken time-reversal symmetry (class IV)}

Now we consider example of a material from class IV.  We will consider ferromagnetic bcc-iron, as it is one of the simplest materials in this class.  We computed the force-constant (dynamical) matrix using the density
functional perturbation theory, as implemented in the Quantum-ESPRESSO computer package,\cite{espresso} within the PBE approximation.\cite{Perdew_1996}  We compute dynamical
matrices on a regular $4 \times 4 \times 4$ grid of q-points in the conventional unit cell with two Fe atoms per cell (this would correspond to effectively $5 \times 5 \times 5$ grid of q-points in the primitive unit cell).

Since bcc-iron has inversion symmetry the phonon angular momentum has to arise from the velocity-force constant matrix $G^{\alpha \beta}_{ij}$, as discussed earlier.  We evaluate interband part 
 of $G^{\alpha \beta}_{ij}$ starting from Eq.~\ref{eq:gij}.  Since this expression includes only the occupied states, we can express $G^{\alpha \beta}_{ij}$ in terms of one-electron orbitals $\phi_{{\bm k} n}$ and occupations $f_{{\bm k}n}$,
\begin{align}
G^{\alpha \beta}_{ij} = 2 \hbar \frac{1}{N_k} \sum_{\bm k} {\rm Im} \sum_n f_{{\bm k}n} \bigg\langle \frac{\partial \phi_{{\bm k} n}}{\partial x^{\alpha}_i} \bigg|
\frac{\partial \phi_{{\bm k}n}}{\partial x^{\beta}_j} \bigg\rangle.
\label{eq:gij2}
\end{align}
We compute the real-space Berry curvature by finite-difference approach.  Let us denote with $\phi_{{\bm k}n}$ a one-electron orbital for a system in which all atoms are at ground state locations and $\phi^{i \alpha}_{{\bm k}n}$ is the one where $i$-th atom is displaced by $\Delta$ in Cartesian direction $\alpha$. The expression for real-space Berry curvature then becomes,
\begin{align}
G^{\alpha \beta}_{ij} = \frac{\hbar}{\Delta^2} \frac{1}{N_k} \sum_{\bm k} {\rm Im} \sum_{mno} 
\Big[
&\left( f_{{\bm k}m}^{1/6} \langle \phi_{{\bm k}m}           | \phi^{i \alpha}_{{\bm k}n} \rangle  f_{{\bm k}n}^{1/6} \right) \notag \\
&\left( f_{{\bm k}n}^{1/6} \langle \phi^{i \alpha}_{{\bm k}n} | \phi^{j \beta }_{{\bm k}o} \rangle  f_{{\bm k}o}^{1/6} \right) \notag \\
&\left( f_{{\bm k}o}^{1/6} \langle \phi^{j \beta }_{{\bm k}o} | \phi_{{\bm k}m}            \rangle  f_{{\bm k}p}^{1/6} \right)
\Big].
\label{eq:gij3}
\end{align}
Clearly, this expression will revert to Eq.~\ref{eq:gij2} in the limit of occupations at zero temperature (in the zero-temperature limit, one can disregard exponents on occupation factors in Eq.~\ref{eq:gij2}). However, this expression has an advantage that unlike Eq.~\ref{eq:gij2} it does not assume that states can be labeled with the consistent band label as atoms are displaced.  In Eq.~\ref{eq:gij3} the sum is done over all states, and since it is manifestly gauge invariant, it therefore doesn't depend on the labeling of bands.

We computed Eq.~\ref{eq:gij3} from first-principles.  We first compute fully relativistic (including spin-orbit) ground state wavefunctions ($\phi_{{\bm k}n}$) of bulk bcc iron in a ferromagnetic state.  Next we construct a $2 \times 2 \times 2$ supercell of the conventional unit-cell (this supercell contains 16 Fe atoms, as there are two atoms of Fe per one conventional unit-cell).  Magnetic moment is set to point along the $z$-axis, which is the easy-axis of the magnetic anisotropy. Next, we rigidly displace one of the Fe atoms in the supercell (labelled with $i=1$) along the $\alpha=x$ direction and repeat the calculation of the electron orbitals (we label these as $\phi^{i=1, \alpha=x}_{{\bm k}n}$). We repeat the same calculation for a displacement of atom along the $\alpha=y$ direction.  There is no need to displace atoms in the $z$ direction, as $G^{\alpha \beta}_{ij} = 0$ by symmetry when either $\alpha$ or $\beta$ are $z$ (we also confirmed this by a direct calculation). The magnitude of the atomic displacements in the $x$ or $y$ direction is $6 \cdot 10^{-3}$~\AA.  There is no need to compute wavefunctions for the displacements of the remaining $16-1=15$ atoms in the supercell, as those can be obtained trivially by real-space translation of wavefunction inside the super-cell.

\begin{figure}[!t]
\centering
\includegraphics[width=3.4in]{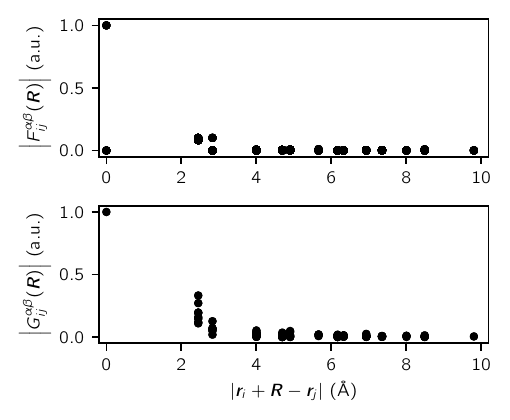}
\caption{\label{fig:decay} Norm of the Fourier transform of force-constant matrix $F^{\alpha \beta}_{ij}$ and velocity-constant matrix $G^{\alpha \beta}_{ij}$ as a function of distance between atoms.}
\end{figure}

Next we obtain the needed $G^{\alpha \beta}_{ij}$ elements by directly computing the overlaps between the wavefunctions as given in Eq.~\ref{eq:gij3}.  To get better spatial resolution of $G^{\alpha \beta}_{ij}$ we also perform Wannier interpolation\cite{Marzari2012} to an effective $4 \times 4 \times 4$ supercell (containing in total 128 Fe atoms).  Figure~\ref{fig:decay} shows that both force-constant matrix $F_{ij}^{\alpha \beta}$ and the velocity-force matrix $G_{ij}^{\alpha \beta}$ decay very quickly in the real-space.

\begin{figure}[!t]
\centering
\includegraphics[width=3.4in]{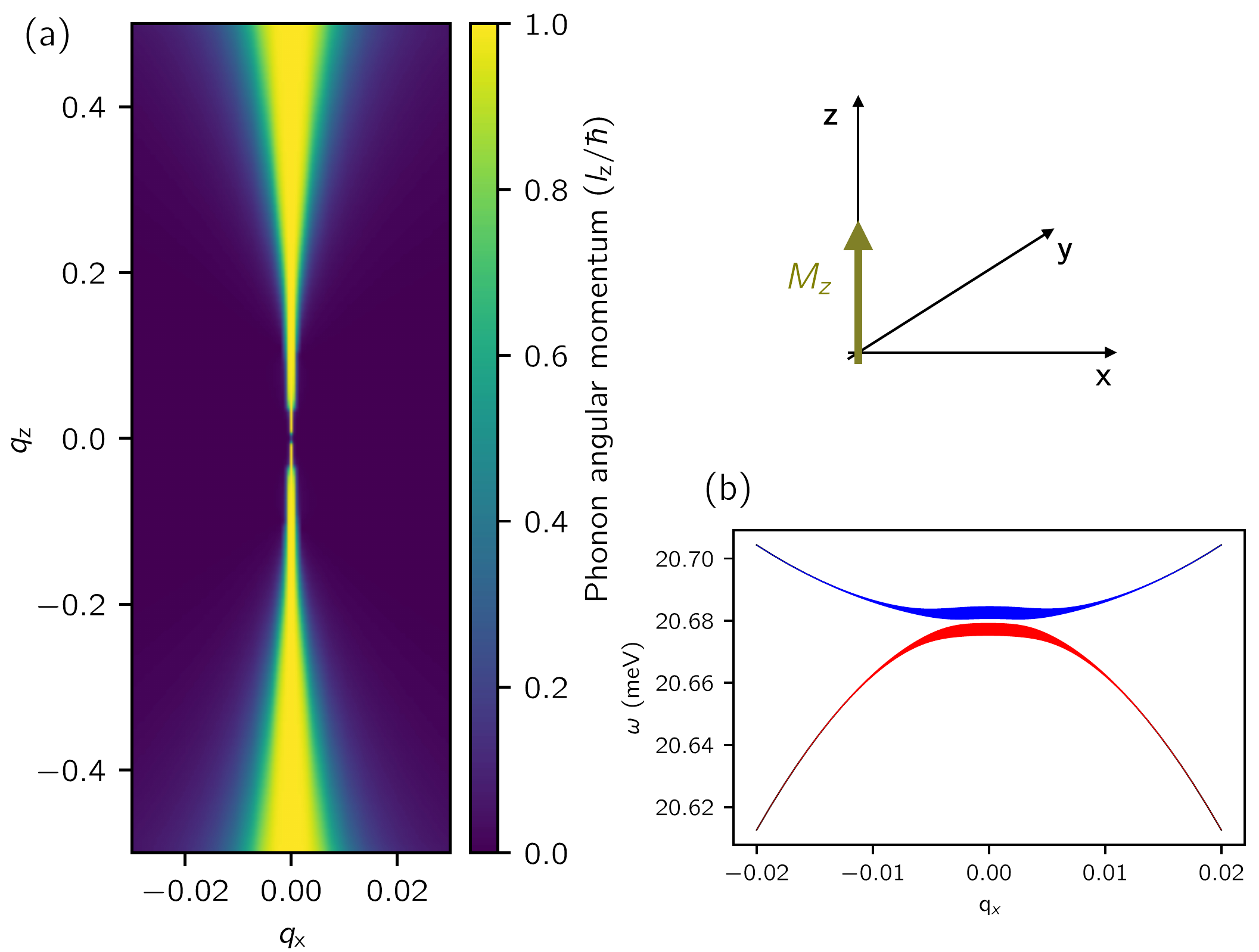}
\caption{\label{fig:fe} Phonon angular momentum of ferromagnetic bcc-iron (class IV).  Phonon wavevector is defined in the conventional unit cell (containing two iron atoms). Phonon wavevector component $q_y$ is set to zero.  (a) Yellow color indicates regions of the Brillouin zone in which two nearly degenerate transverse acoustic phonons are circularly polarized ($l=\pm \hbar$) due to the velocity-force term $G^{\alpha \beta}_{ij}$.  Phonons in purple regions have a negligible amount of phonon angular momentum.  The magnetization of iron is pointing along the $z$ axis.  Panel (b) shows the cut through the phonon Brillouin zone (defined by $q_y=0$ and $q_z=0.4$) using the same coloring convention for the phonon angular momentum as in Fig.~\ref{fig:wc}.}
\end{figure}

Results for the phonon dispersion in iron, including the effects of the velocity-force constant matrix $G^{\alpha \beta}_{ij}$, are shown in Fig.~\ref{fig:fe}.  The addition of the velocity-force constant term introduces a small gap-opening between the two transverse acoustic branches. In the vicinity of the gap-opening, the phonons are fully circularly polarized.  (We computed phonon angular momentum using Eq.~\ref{eq:ldef} which is valid in class IV and V as long as $G^{\alpha \beta}_{ij}$ is small.\footnote{Reference~\onlinecite{Saparov2021} derives phonon angular momentum by carefully considering the difference between the kinetic and canonical momentum.  In classes IV and V this
distinction lead to a small correction to the phonon angular
momentum, proportional to $(\omega - \omega_0) / \omega_0$.  Here
$\omega_0$ is the phonon frequency when all $G_{ij}^{\alpha \beta}$
in Eq.~\ref{eq:eigval} are set to zero.})
These regions with non-zero phonon angular momentum are indicated with a yellow color in Fig.~\ref{fig:fe}. Since without the velocity-force term the transverse phonons are exactly degenerate along $q_x=q_y=0$ line the velocity-force constant term introduces fully quantized angular momentum in the vicinity of the line regardless of its strength.  Of course, with stronger spin-orbit coupling we expect that the induced velocity-force will be larger.  For example, the phonon gap opening in CeF$_3$,\cite{Shaack1975} that likely originates from the same microscopic mechanism, as discussed in Ref.~\onlinecite{Zhang2014}, is about 25~cm$^{-1}$ which is 6~\% of the phonon frequency. Recent theoretical work on a different material, ferromagnetic CrI$_3$, finds significantly smaller phonon splitting, since the relevant phonon frequency is much lower than the acoustic magnon frequency at $\Gamma$.\cite{bonini2022}

\section{Conclusion}

We performed symmetry analysis to understand which materials can or can't have phonons with angular momentum at generic non-symmetric parts of the Brillouin zone (away from high-symmetry point, line, or plane).  All materials fall into one of the five classes, depending on whether $\cal P$, $\cal T$, and ${\cal P T}$ are symmetries or not.  Here, spatial inversion symmetry is denoted as $\cal P$ while time-reversal is denoted as $\cal T$.  The time-reversal breaking in the phonon dynamics does not occur in the force-constant matrices. Instead one must go to the next term in the expansion, the velocity-force constant matrix $G^{\alpha \beta}_{ij}$.  The $G^{\alpha \beta}_{ij}$ measures force on atom $i$ induced by velocity, not displacement, of atom $j$.  These effects will be relevant not only for phonon angular momentum in class IV, but also for any other effect that depends on time-reversal breaking in phonons, such as phonon Hall effect, magnetic moment of a phonon, and Einstein de-Haas effect.

\acknowledgements{This work was supported by grant NSF DMR-1848074.  I acknowledge discussions with Richard Wilson, Massimiliano Stengel, and Cyrus Dreyer.}

\bibliography{pap}

\end{document}